\documentclass [prb,twocolumn,tightenlines,amssymb,showpacs]{revtex4}
\usepackage{amsmath}
\usepackage {graphicx}
\begin{document}

\title{Temperature independent diffuse scattering and elastic lattice deformations in relaxor PbMg$_{1/3}$Nb$_{2/3}$O$_3$}

\author{R. G. Burkovsky}\email {rg.burkovsky@mail.ioffe.ru}
\affiliation{St.Petersburg State Politekhnical University, 29 Politekhnicheskaya, 195251, St.-Petersburg, Russia}

\author{S. B. Vakhrushev}
\affiliation{Ioffe Phys.-Tech. Institute, 26 Politekhnicheskaya, 194021, St.-Petersburg, Russia}
\affiliation{St.Petersburg State Politekhnical University, 29 Politekhnicheskaya, 195251, St.-Petersburg, Russia}

\author{K. Hirota}
\affiliation{Structural Physics in Extreme Conditions Group, Department of Earth and Space Science, Graduate School of Science, Osaka University, Osaka 560-0043, Japan}

\author{M. Matsuura}
\affiliation{Structural Physics in Extreme Conditions Group, Department of Earth and Space Science, Graduate School of Science, Osaka University, Osaka 560-0043, Japan}

\date{\today}

\begin{abstract}

The results of diffuse neutron scattering experiment on PbMg$_{1/3}$Nb$_{2/3}$O$_3$ single crystal above the Burns temperature are reported. It is shown that the high temperature elastic diffuse component is highly anisotropic in low-symmetry Brillouin zones and this anisotropy can be described using Huang scattering formalism assuming that the scattering originates from mesoscopic lattice deformations due to elastic defects. The qualitative agreement between this model and the experimental data is achieved with simple isotropic defects. It is demonstrated that weak satellite maxima near the Bragg reflections can be interpreted as the finite resolution effect.

\end{abstract}

\pacs{61.72.Dd, 77.84.Dy}

\maketitle
\bigskip

\section{INTRODUCTION}
Ferroelectrics with diffuse phase transition were first discovered by Smolensky~et~al. \cite{Smolensky:1958} more than fifty years ago. Now they are usually referenced as relaxor ferroelectrics \cite{Cross:1987}. These materials demonstrate exceptional piezoelectric capabilities \cite{Park:1997} and very high dielectric constants in a broad temperature range making them promising materials for applications in electronic and electromechanical devices. The most distinguished features of relaxors are broad frequency dependent maximum of dielectric permittivity $\varepsilon(T)$ \cite{Colla:1992:Low-frequency, Bovtun:2006} and the nonergodicity of low-temperature phase \cite{Colla:1992:Long-time}. Above the characteristic Burns temperature $\mathit{T_d}$ relaxors behave similarly to normal displacive ferroelectrics - $\varepsilon(T)$ follows Curie-Weiss law \cite{Viehland:1992} and the soft mode is observed \cite{Naberezhnov:1999}. 
Typical relaxors are perovskite-like mixed crystals with one or two sublattices randomly occupied by nonisovalent ions. The most studied relaxor compound is lead magnoniobate PbMg$_{1/3}$Nb$_{2/3}$O$_3$ (PMN) having B-sublattice populated by Mg$^{2+}$ and Nb$^{5+}$ in proportion 1/2. Diffraction studies show that without application of strong electric field PMN crystals retain cubic structure at least down to to 5 K and no ferroelectric phase transition takes place \cite{Bonneau:1989,deMathan:1991}. However it is clear that at least below $\mathit{T_d}\approx 650$ K local structural distortions develop \cite{Burns:1983, Dmowski:2008}. 
These distortions are reflected by quasi-elastic diffuse scattering (DS) that appears below $\mathit{T_d}$ in the vicinity of the Bragg reflections \cite{Xu:2004:Neutron, Naberezhnov:1999}. The scattering is highly anisotropic in reciprocal space and has different shapes in different Brillouin zones \cite{Xu:2004:Neutron}. On cooling down below $\mathit{T_d}$ the intensity of DS monotonically increases and at cryogenic temperatures becomes comparable with the intensity of Bragg reflections. There were proposed numerous interpretations of DS in relaxors including scattering on highly anisotropic optic \cite{You:1997} and acoustic \cite{Yamada:2002} phonon modes, scattering on static correlated ionic displacements in polar nanoregions \cite{Xu:2004:Three} and scattering on static mesoscopic lattice deformations caused by polar defects \cite{Vakhrushev:2005}. 
The Monte-Carlo \cite{Welberry:2006,Pasciak:2007} and molecualr dynamics \cite{Ganesh:2010} simulations were also productively emoloyed in the recent years to describe the spatial distribution of the DS in relaxors. 
The temperature-dependent DS can be considered as one of the relaxor footprints but despite extensive studies its microscopic origin is not yet clear. 
 
\begin{figure}
\includegraphics [width=\columnwidth,clip=true, trim=0mm 0mm 0mm 0mm] {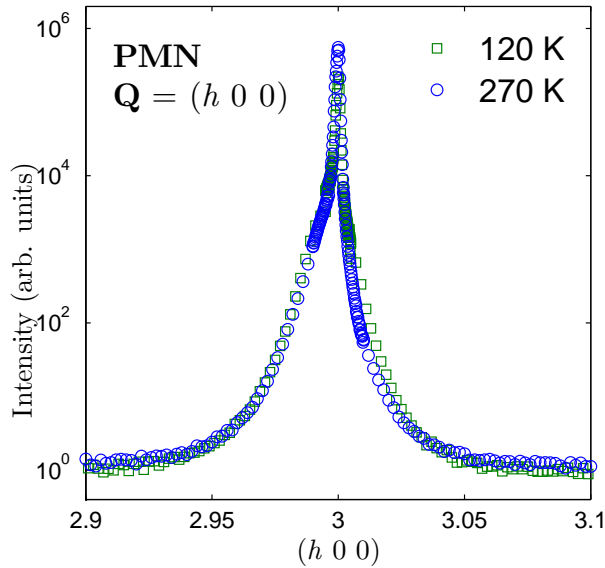}
\caption{\label{fig_scans1996} (Color online) The longitudinal (radial) scans of synchrotron X-ray diffuse scattering in PMN in the vicinity of (300) Bragg reflection for $T$=120 K (squares) and $T$=270 K (circles) \cite{Vakhrushev:1996}. }
\end{figure}

Along with temperature-dependent DS a nearly temperature-independent components were observed in PMN and related compounds \cite{Vakhrushev:1996, Hiraka:2004, Stock:2006}. First observation was reported in Ref.\onlinecite{Vakhrushev:1996} where the DS was studied in PMN using high-resolution synchrotron X-ray scattering. It was emphasized that the longitudinal (or radial) scans contain a component that does not change at least in 80~K~$<~T~<$~270 K range (Fig. \ref{fig_scans1996}). This component was interpreted by the authors as Huang scattering due to elastic lattice deformations.
In 2004 using neutron scattering Hiraka~et~al. \cite{Hiraka:2004} clearly observed weak temperature independent component of DS in (100) and (110) BZs both above and below $\mathit{T_d}$ and attributed it to the correlated displacements of Pb$^{2+}$ ions due to short-range chemical order which is believed to be temperature insensitive up to $T\approx 1000$ K. This interpretation was supported in further study by Gehring~et~al. \cite{Gehring:2009} but opposed in molecular dynamics study of relaxor PSN \cite{Ganesh:2010} where the radial DS was attributed to the chemical disorder.
To clarify the nature of the temperature independent radial component of DS in lead magnoniobate we extended the measurements of neutron diffuse scattering to BZs of low symmetry. In this contribution we show that our data and the data from Ref.\onlinecite{Hiraka:2004} can be adequately described in terms of Huang scattering due to inhomogenous lattice deformations which are definitely expected in mixed crystals.

\section{EXPERIMENTAL DETAILS}

\begin{figure*}
\includegraphics [width=\textwidth,clip=true, trim=15mm 15mm 15mm 0mm] {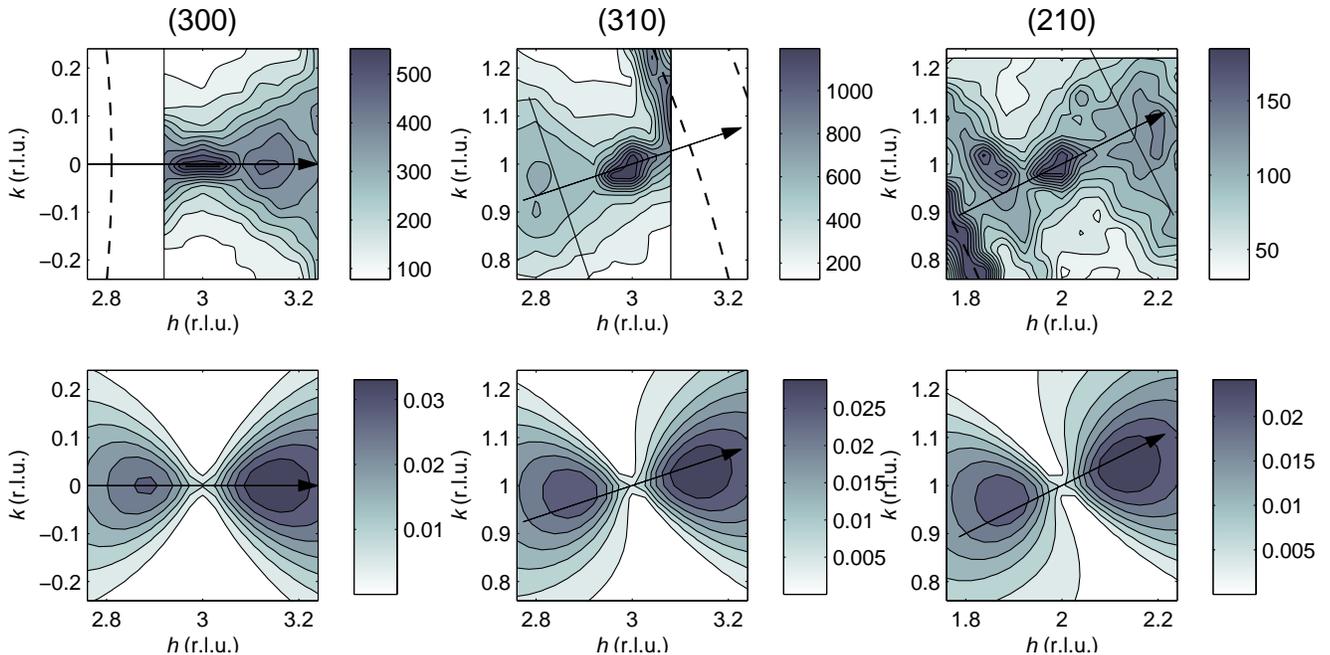}
\caption{\label{fig_2dmaps} (Color online) Isointensity contours of diffuse scattering in PMN at $T$=650 K in three BZs (top) and corresponding isointensity contours produced by model calculations described in text (bottom). The arrows through the zone centers denote the direction of reciprocal lattice vector {\Large $\mathbf{\tau}$}. The positions of Al powder coils are marked by dashed lines. The straight solid lines perpendicula to vector {\Large $\mathbf{\tau}$} mark the positions in which the slices depicted on Fig. \ref{fig_kososlices} are made. }
\end{figure*}

PMN single crystal was supplied by the Institute of Physics, Southern Federal University (Rostov-on-Don, Russia). The measurements were carried out at 5G-PONTA three-axis spectrometer installed in the reactor hall of the JRR-3M reactor (Tokai, Japan). The sample was wrapped in the Al foil and mounted in the furnace with (001) axis vertical. The experimental setup allowed heating up to 650K . 
Two-dimensional distributions of the diffuse scattering intensity were measured at $\mathit{T}$ = 650 K in (300), (310) and (210) Brillouin zones in quasielastic mode with fixed wavevector $K_i=K_f=3.84$\AA$^{-1}$ (30.5 meV). Energy resolution was of the order of 1 meV and Q-resolution (FWHM) was $\Delta Q_L=0.06$\AA$^{-1},\ \Delta Q_T=0.02$\AA$^{-1},\ \Delta Q_V=0.4$\AA$^{-1}$ for the component along the scattering vector, in-plane transverse component and out-of-plane (vertical) transverse component respectively. 
Due to strong contamination by Al powder scattettering in (300) and (310) Brillouin zones some parts of 2-d maps on Fig. \ref{fig_2dmaps} are omitted. In order to clearly distinguish between Al powder scattering and PMN diffuse scattering the positions of Al powder coils are marked by dashed lines.

\section{RESULTS AND MODELLING}
\begin{figure}
\includegraphics [width=\columnwidth,clip=true, trim=0mm 0mm 0mm 0mm] {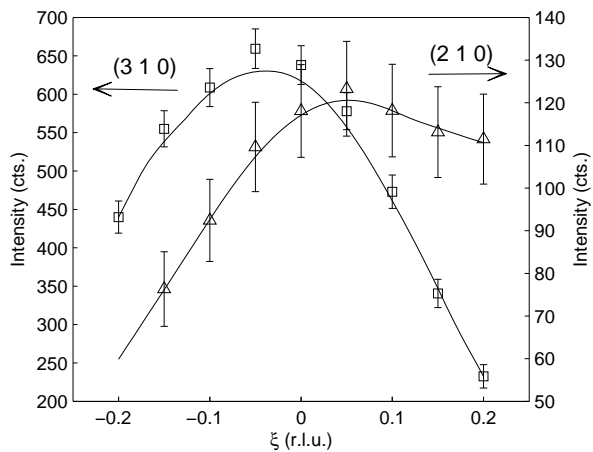}
\caption{\label{fig_kososlices} The 1-d slices of 2-d diffuse scattering maps performed along the lines perpendicular to the {\Large $\mathbf{\tau}$} vector (see text for the details). These lines are shown in solid on Fig. \ref{fig_2dmaps}.}
\end{figure}

In general the results of our experiment are consistent with the previous ones \cite{Hiraka:2004}: in high-symmetry (300) Brillouin zone the scattering is oriented along the direction of reciprocal lattice vector {\Large $\mathbf{\tau}$}  which is denoted by the arrows on Fig. \ref{fig_2dmaps}, and the longitudinal profile contains weak maximum on approximately 0.15 r.l.u. from the zone center. In zones of lower symmetry the situation is somewhat different - the dumbbell-like diffuse scattering shapes are not purely longitudinal. The areas with maximal intensity are clearly shifted from the line $L_\tau$ that passes through the reciprocal lattice point and is oriented along {\Large $\mathbf{\tau}$}. Even more clearly it is seen on 1-d slices of 2-d maps presented on Fig. \ref{fig_kososlices}. The slices are made along the straight lines in reciprocal space determined by the vector equation $((\mathbf{\xi - \xi_0}) \cdot \mathbf{\tau}) = 0$ where $\mathbf{\xi} = (h, k, 0)$ and the center of the slice $\xi_0$ lies on $L_\tau$. These lines are shown in solid on Fig. \ref{fig_2dmaps}.

We look for the origin of the observed scattering in elastic lattice deformations that are unavoidable in mixed crystals. This kind of scattering was heavily studied in alloys and other systems with imperfect lattice. In the book of M.~A.~Krivoglaz \cite{Krivoglaz:1996} a useful formalizm for calculating of the distribution of static ionic displacements in the lattice due to elastic defects is presented. The characteristics of the spatial distribution of atomic displacements due to elastic deformations are determined by the types of defects and by elastic constants of the crystal. In our analysis we used the most straightforward way of applying this formalism by assuming simple cubic symmetry defects in PMN lattice whose elastic constants are known in advance \cite{Lushnikov:2008}.
The Fourier components of lattice deformation field are described by the expression 
\begin{equation}
\label{eq_displacements}
A_{\alpha i}(\mathbf{q}) = \frac{1}{q} N_d C_{ij}^{-1}(\mathbf{q}) p_{\alpha jl}n_l
\end{equation}
where $N_d$ is the number of defects, $\mathbf{q}$ - reduced wavevector, $\mathbf{n}={\mathbf{q}}/q$. The tensors $C_{ij}(\mathbf{q})=c_{iljm}n_ln_m$ and $p_{\alpha ij}=\sum{c_{ijlm}L_{\alpha lm}}$ are calculated on the basis of elastic constants of the crystal $c_{ijlm}$. The elastic deformation tensor $L_{\alpha jl}$ describes the symmetry and "force" of defect with orientation $\alpha$. 
The intensity of diffuse scattering due to the elastic deformations (\ref{eq_displacements}) is determined by expression
\begin{equation}
\label{eq_intensity}
I_H(\mathbf{Q}) = |F(\mathbf{\tau})|^2 
\sum_{\alpha=1}^{\nu} (\mathbf{Q}\cdot \mathbf{A}_{\alpha}(\mathbf{q}) )^2
\end{equation}
The scattering vector $\mathbf{Q}$ is a sum of reciprocal lattice vector {\Large $\mathbf{\tau}$} and reduced vawevector $\mathbf{q}$, $F(\mathbf{\tau})$ is the elastic structure factor. Summation is made over $\nu$ possible different orientations of defects ($\nu>1$\ if the symmetry of the defects is lower than that of the crystal). 

The equation (\ref{eq_displacements}) describes the deformations of an ideal anisotropic elastic medium with low concentration of defects and predicts an infinite increase of $A_\alpha (\mathbf{q})\mid_{q\rightarrow 0}$. The corresponding intensity of scattering is proportional to $1/q^2$. 
In real crystals one obviously could observe only finite intensities and one of the mechanisms that can substantially smooth the profiles of Huang-type scattering near the ZC is lattice deformation screening. In the case of large concentration of defects the deformation fields produced by individual defects are spatially screened due to the influence defects in the neighborhood. This results in exponential cutoff factor for the deformation fields in the form

 $u(r) \sim \frac{1}{r}e^{-r/r_s}$ 
 
 where $u(r)$ is the strain magnitude and $r_s$ - the screening length \cite{Emel'yanov:2001}. 
In the reciprocal space this fact is reflected by changing from $1/q^2$ dependence to $1/(q^2+q_s^2)$.

We account for this effect by modification of (\ref{eq_displacements}) in the form 
\begin{equation}
\label{eq_screened_displacements}
A_{\alpha i}(\mathbf{q}) = \frac{1}{\sqrt{q^2+q_s^2}} N_d C_{ij}^{-1}(\mathbf{q}) p_{\alpha jl}n_l
\end{equation}
This way the scattering profies transform into Lorentzians with parameter $q_s$ inverse proportional to the screening length $r_s$. The experimental data show no sharp increase of diffuse intensity near the Bragg reflections (Fig. \ref{fig_slice}) and we attribute this fact to the presence of strong deformation screening in PMN. 

\begin{figure}
\includegraphics [width=\columnwidth, clip=true, trim=5mm 5mm 0mm 0mm] {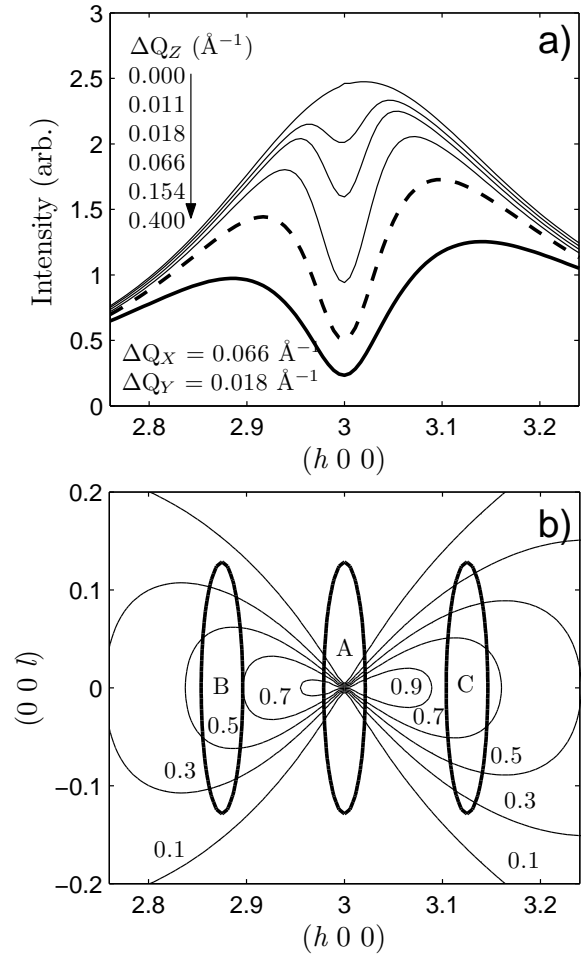}
\caption{\label{fig_conv} 
a) A set of modelled radial scans of diffuse scattering (without Bragg and background contributions) near (3~0~0) reflection with different resolution conditions. The thick line corresponds to the estimated resolution of current experiment, dashed line - to the estimated resolution of experiment from Ref. \onlinecite{Hiraka:2004}. The uppermost line corresponds to the ideal resolution, the second upmost line corresponds to $\Delta Q_Y=\Delta Q_Z=0.011$~\AA$^{-1}$, for other lines the resolution conditions are denoted on the plot. 
b) The projections of resolution ellipsoids (thick lines) for the zone center (A), $q$=-0.125 (B) and $q$=0.125 (C) and isointensity contours of the Huang scattering with parameters described in text. Resolution function and scattering intensity are normalized to unity, the thick lines correspond to the value $I$=0.5.}
\end{figure}

As it was mentioned before we consider the defects of cubic symmetry which are characterized by elastic deformation tensor of diagonal form $L_{\alpha jl}=\delta _{jl}\Lambda$. The defects of this kind expand or compress the surrounding lattice equivalently in all three main crystallographic directions, the character (compression or expansion) is determined by the sign of constant $\Lambda$. 
In case of cubic defects $I_H(\mathbf{Q})$ has a zero-intensity plane that is nearly normal to the vector {\Large $\mathbf{\tau}$} \cite{Krivoglaz:1996} and the intensity near the zone center can be approximated by expression $const \cdot cos(\phi)/q$ where $\phi$ is the angle between {\Large $\mathbf{\tau}$} and $\mathbf{q}$. This expression results in appearance of the thin waist near the zone center outside which the intensity is nearly zero (area A on Fig. \ref{fig_conv}(b)). 
Calculated 2-D distributions of DS intensity convoluted with experimental resolution function are presented in the bottom row of Fig. \ref{fig_2dmaps}. 
One can clearly see that the three main peculiarities - the character of anisotropy, the presence of the waist and the satellite maxima are well reproduced by the model.

\section{SATELLITE PEAKS}

In all studied Brillouin zones the diffuse scattering intensity contains maxima at finite $q$ values. In high symmetry zones the peak positions lie on the line going through the corresponding reciprocal lattice points. In zones of lower symmetry these positions are somehow shifted from that line. Equations (\ref{eq_screened_displacements}) and (\ref{eq_intensity}) describe the intensity that monotonically decreases with increasing $q$ value. In this section we show how finite experimental resolution can affect the scattering profiles resulting in the appearance of the observed peaks.  
In the case of elastic scattering the resolution function (RF) can be approximated by 3-D gaussian \cite{Cooper:1967}:
\begin{equation}
R(\mathbf{q}) =R_0 \cdot  e^{ -\frac{1}{2} (   
(q-q_0)_{\parallel}^2 / \sigma_{\parallel}^2   
+  (q-q_0)_{\perp}^2 / \sigma_{\perp}^2
+ (q-q_0)_V^2 / \sigma_{V}^2
)}               
\end{equation}
Here the values $(q-q_0)_\parallel$ ... $(q-q_0)_V$ are the components of vector $(\mathbf{q-q_0})$ along the vector {\Large $\mathbf{\tau}$}, along the in-plane and out-of-plane perpendiculars correspondingly. In our experiment the in-plane components of the RF were quite sharp as it can be deduced from the images of the Bragg reflections on Fig. \ref{fig_2dmaps}(a). In order to provide high neutron flux the vertical collimations were highly relaxed which resulted in more broad resolution in out-of-plane (vertical) direction. This is illustrated on Fig. \ref{fig_conv}(b) by the $R(\mathbf{q})=R_0/2$ resolution ellipsoids stretched in (0 0 $l$) direction. 

\begin{figure}
\includegraphics [width=\columnwidth,clip=] {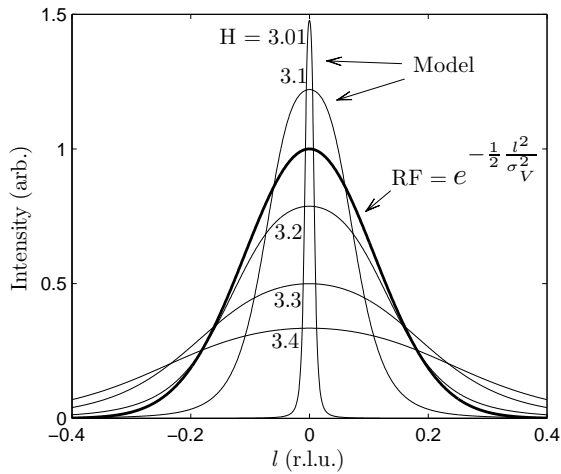}
\caption{\label{fig_crossover}
The slices of 3-D model scattering intensity in (3 0 0) Brillouin zone determined by equations (\ref{eq_screened_displacements}) and (\ref{eq_intensity}) along the vertical $l$ axis with fixed value of $k$=0 and different values of $h$ described on the graph. Bold line represents the normalized resolution function.}
\end{figure}

The figures \ref{fig_conv}(b) (2-D contour plots) and \ref{fig_crossover} (1-D slices) represent the same distribution of the intensity calculated using equations (\ref{eq_screened_displacements}) and (\ref{eq_intensity}) in (3 0 0) Brilluin zone together with the resolution function. One can see from these figures that (a) the intensity decreases with increasing $q$ value and (b) the distribution of intensity along $q_\perp$ directions is very narrow for small $h$ values and is broadening on moving from the zone center along ($h$ 0 0). The first factor leads to the decrease of the detector count rate with increasing $h$ value. The second factor on the other hand leads to its increase due to increasing of overlapping between the intensity distribution and the RF. Near the zone center the second factor is dominant which results in the increase of the count rate. This trend changes near $h$=0.15 r.l.u. where the widths of the intensity distribution and the RF become comparable and the impact of the second factor diminishes. On further increase of $h$ the detector count rate follows the decrease of the unconvoluted intensity (Fig. \ref{fig_conv}(a)). The crossover between these two regimes is marked by satellite maxima at $q$=0.15 r.l.u.

\begin{figure}
\includegraphics [width=\columnwidth,clip=] {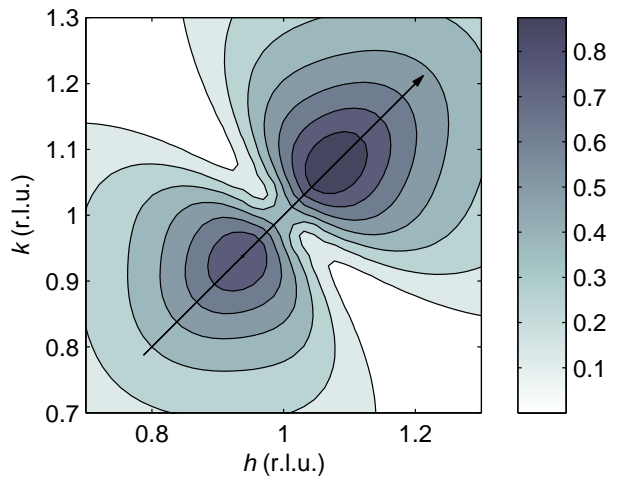}
\caption{\label{fig_hiraka} (Color online) Calculated diffuse scattering intensity in (110)  Brillouin zone. Corresponding experimental data were reported in Ref.\onlinecite{Hiraka:2004}.}
\end{figure}

The maxima do shift towards larger $q$ values with increasing $\sigma$ values. On Fig.\ref{fig_conv}(a) the profiles for different $\sigma_V$ values are presented including the conditions of our experiment (bold solid line) and experiment by Hiraka et al. \cite{Hiraka:2004} (dashed line). Their vertical resolution should be about 2.6 times better than our due to use of cold neutrons with $E_i=E_f=4.5$ meV. The resolution effect in this case is is still strong enough to produce the observed satellite maxima. The 2-D map calculated with these experimental conditions for (1 1 0) BZ (Fig. \ref{fig_hiraka}) is in very good qualitative agreement with the experiment. 

\begin{figure}
\includegraphics [width=\columnwidth, clip=true, trim=0mm 0mm 0mm 0mm] {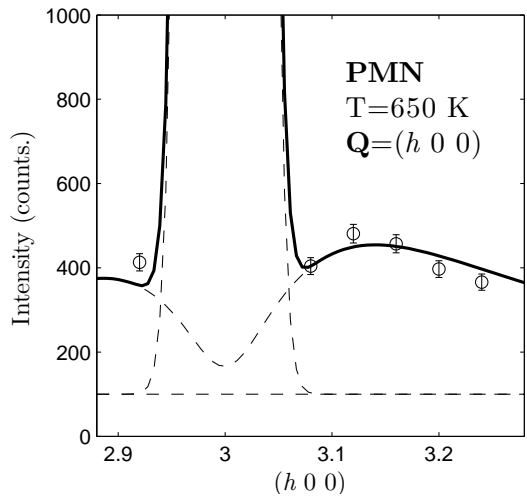}
\caption{\label{fig_slice} The longitudinal (radial) scan of diffuse scattering in PMN at T=650 K in the vicinity of (3~0~0) Bragg reflection (circles), model fit (solid line) and the decomposition of fit into the Bragg, diffuse and background components (dashed lines).}
\end{figure}

In all calculations presented in this paper we used the reciprocal screening length $q_s=0.2$ which we found to be optimal for the description of the scattering profiles (Fig.\ref{fig_slice}).

\section{DISCUSSION}
We should discuss the microscopic characteristics of PMN that can be responsible for the observed diffuse scattering. Relaxors are mixed crystals and are definitely expected to suffer from elastic deformations due to the intrinsic inhomogenity. Indeed the deformations modelled on the basis of known in advance elastic constants and the most simple cubic symmetry defects are shown to produce the scattering with the same shapes as obtained in the experiment. 
The role of defects most probably play the inhomogenities of chemical composition. It is naturally to expect that the relative density of Mg$^{2+}$ and Nb$^{5+}$ ions is weakly fluctuating on the nanoscale. The areas characterized by excess of one type of these ions are expected to have different from average unit cell volume due to difference in effective ionic radii (0.72 \AA for Mg$^{2+}$ and 0.64 \AA for Nb$^{5+}$)\cite{Shannon:1976} and thus produce expansion or compression of the surrounding lattice. 
The areas with excess of one or another type of ions most probably should be considered not as well-localized entities but as entities smoothly passing into each other. The cross-overs between regions with excesses of different ions take place on the nanoscale so it is naturally to assume that the concentration of defects is large enough to produce deformation screening needed to suppress the sharp increase of diffuse intensity near the zone centers. Possibly the chemically ordered regions also contribute to the lattice deformations due to the different from the disordered matrix unit cell volume.

Also we address the problem of satellite maxima near the Bragg reflections. Typically the satellites are created when the new periodicity longer than the unit cell is created. The positions of the satellites are determined by the ratio of the "new" and "old" periods . In the case of high-temperature diffuse scattering in PMN the positions of maxima vary between different BZs but remain on the nearly the same distance from BZs center. If we assume that the peaks originate from short range correlations of ionic displacements as it was suggested in Ref. \onlinecite{Hiraka:2004} it seems to agree well with anomalously large width of the peaks but it also seems very difficult to construct a quantitative model predicting the positions of the peaks. 
The model proposed here together with the resolution effect provide on the other hand a simple way of satellite peak description without any additional assumptions except the lattice deformation screening which by itself is quite expectable in relaxors.


\section{CONCLUSION}
In this contribution we report the results of the detailed study of the high-temperature diffuse scattering in low-symmetry Brillouin zones in PMN and propose a model that accounts well for both the discovered scattering anisotropy and the presence of satellite maxima. 
The assumptions on which the model is based are simple and consistent with the data obtained earlier. 
To clarify the microscopic origin of the elastic lattice deformations additional measurements are necessary.

\begin{acknowledgments} 
It is a pleasure to acknowledge T.~Egami, B.~Burton, P.~Gehring, R.~Cohen, S.~Prosandeev and V.~Sakhnenko for many useful discussions and various suggestions. The PMN single crystals were provided by the Institute of Physics Rostov-on-Don University. SBV acknowledges the support of Institute of Solid State Physics, University of Tokyo. Work at the Ioffe Institute was supported by the RFBR (grants 08-02-00908-a and 06-02-90088-NSF-a ) and RAS Program "Neutron study of structure and fundamental properties of Matter". The work at St.-Petersburg State Polytechnical University was supported by Federal Program "Scientific and educational staff of innovative Russia" for 2009-2013 years and by grant of the St.-Petersburg government.
\end{acknowledgments}

\bibliography{mybibliohtds}

\end{document}